\documentstyle[twocolumn,seceq,epsfig]{jpsj}

\def\runtitle{A scenario for the dynamics in the small entropy production limit}
\def\runauthor{Leticia {\sc Cugliandolo} and Jorge {\sc Kurchan}}

\title
{
A scenario for the dynamics in the small entropy production limit
}

\author
{ 
Leticia {\sc Cugliandolo}
and Jorge {\sc Kurchan}$^{1}$
}

\inst
{
Laboratoire de Physique Th\'eorique, \'Ecole Normale Sup\'erieure, 
24 rue Lhomond, 75231 Paris Cedex 05, France\\
Laboratoire de Physique Th\'eorique et Hautes \'Energies, Jussieu, 
4, Place Jussieu, 75005 Paris, France\\
$^1$PMMH, \'Ecole Sup\'erieure de Physique et Chimie Industrielles, 10 rue Vauquelin, 75231 Paris Cedex 05, France
}

\recdate
{
\today
}

\abst
{
We present a scenario for the nonequilibrium dynamics 
in the limit of small entropy production. 
We discuss (i) the appearence of different time-scales, 
(ii) the modification of the fluctuation-dissipation theorem
and its relation to effective temperatures and partial equilibrations 
and (iii) the validity of Onsager reciprocity relations. 
We distinguish these properties by their reaction to infinitesimal 
perturbations. We recall that one can easily change the 
time dependence of observables by applying an infinitesimal force while
time-reparametrization invariant features 
remain unchanged under the same perturbations. With the aim of better 
understanding these properties,  we consider 
the effect of several
baths with different temperatures and time-scales on the dynamics. 
This is done in two ways: numerically, by using a especially developed 
Monte Carlo algorithm 
that mimics the coupling to
multiple baths; analytically, by computing 
the time-dependent probability density of  
simple systems in contact with multiple baths.
We finally argue that  these features are related to 
supersymmetry, the reparametrization invariance of the slow dynamics and its 
spontaneous breaking. This scenario is consistent 
within any perturbative resummation scheme. A brief version of this 
article appeared in Ref.~\cite{Cuku-stat}
}

\kword
{
nonequilibrium dynamics, effective temperatures
}

\begin{document}
\sloppy
\maketitle

\section{Introduction}

In many  dynamical systems, there is a quantity that can
be naturally interpreted as the entropy production.\cite{Kubo}
It depends on the time-dependent probability distribution
and vanishes only when the system reaches the Gibbs-Boltzmann (GB) measure.
The limit of small, though non-vanishing, 
entropy production (SEP) is achieved in a relaxing system 
at long times,  and in  stationary systems driven by external forces (or by different 
thermal baths) when the external power input tends to zero.

A {\em finite} system in which all forces  derive from a potential 
approaches, for any reasonable dynamics and in   
the long-time  limit, the GB distribution. 
In contrast, an {\it extended} system
might stay far from equilibrium even at very long
times, producing entropy at a small rate. In this SEP limit 
the probability distribution 
{\it is not} a small perturbation around the GB equilibrium one. 

There are several examples of such nonequilibrium systems.
An infinitely large system  with growing domains 
has, at any finite time, properties that are  
dominated by the domain walls\cite{Bray}
and is out of equilibrium.
Similarly, glassy systems `age': their responses and correlations depend not
only on time-differences but also on the time elapsed since their preparation.
\cite{aging_polymers,aging_spinglasses,aging_orient,aging_glycerol,aging_colloids,megen,review,Bouchaud} 
Both  facts are   taken into account neither 
by the GB distribution nor by a simple perturbation around it.
 
On the other hand, 
a system that would otherwise age might be kept 
in a stationary yet far from 
equilibrium regime of sustained (relative) youth
by pumping a small quantity of
energy.\cite{Crisanti,Cukulepe,Onuki,Jorge,berthier}
 Such thing would happen if domains are kept on average
 from growing by  small, nonconservative  forces \cite{Onuki}.
 While the experimental situation for glasses is less clear in this respect, 
simple models suggest that  if one 
subjects a system below the glass temperature
to  shearing forces, 
the viscosity may stop growing at a large, 
driving-power dependent value.\cite{Chaveteau} 
This situation is reminiscent of what happens 
in granular flow.\cite{Nagel}
In these cases the systems are explicitly kept far from equilibrium by the 
 `non-potential' forces.

  If the SEP limit is not the GB measure, 
is there any generic, model independent statement we can make about it? 
The purpose of this paper is to discuss several aspects of a  scenario
with partial equilibrations at separate time-scales\cite{Cuku-stat}  inspired
by `mean-field' glass theory.\cite{review} 
 Although we cannot show that this scenario is the most general one,
we can show that it is consistent in any dimension.
It leads to definite predictions, some of which have already been tested. 
and some  new ones that, to a limited extent, we test here.

The  discussion  will be aimed at  extended systems 
satisfying reasonable spatial homogeneity properties. We may easily 
construct counterexamples of the present scenario in models with disconnected
or almost disconnected parts, in which microscopic 
couplings scale with the system size, etc.   
Throughout this paper whenever a quantity is declared to be `small', it must be understood 
with the zero entropy production limit taken first, and this itself  taken after the infinite size limit.

\section{Formalism}

Consider a system described by the variables 
$\phi_i$, $i=1,\dots,N$, that we encode in a
vector ${\vec \phi}$, with energy $E({\vec \phi})$.
A nonconservative force ${\vec f}$, scaled with a parameter $D$, 
acts on ${\vec \phi}$.
If the system is set in contact with a {\it white} thermal bath, 
the uncorrelated Gaussian noises $\eta_i$ have zero mean and variance $2\Gamma_0 T$.
$\Gamma_0$ is the friction coefficient.
The equations of motion are
\begin{equation}
-m_i {\ddot \phi}_i - \frac{\partial E({\vec \phi})}{\partial \phi_i}  
+ D f_i({\vec \phi}) =
 \Gamma_0 {\dot \phi}_i - \eta_i
\label{motion}
\end{equation} 
and the masses $m_i$ can be zero in particular. 

If the system is coupled to a 
{\it coloured} Gaussian bath,\cite{Hanggi}
the noise and friction terms are
\begin{equation}
\int_{t_o}^t  ds \; \nu(t,s) {\dot \phi}_i (s) - \rho_i  
\label{baba}
\end{equation}
with
$
\langle \rho_i(t)\rho_j(t_w) \rangle =
T^* \, \delta_{ij} \, \nu(t,t_w) 
$
and $T^*$ the temperature of the bath.
The presence of the same kernel $\nu(t,s)$ in the noise-noise correlation 
and in the friction term is a consequence of having taken a bath in equilibrium. 

We define the  correlation functions and the linear response of the variable  
$\phi_i$ to a kick applied to $\phi_j$ at $t_w$: 
\begin{eqnarray*}
 C_{ij}(t,t_w)  =\langle \phi_i(t) \phi_j(t_w) \rangle 
\;, 
R_{ij}(t,t_w) = 
\left. \frac{\delta \langle \phi_i(t) \rangle }{\delta h_j(t_w)}
\right|_{h=0}
\; .
\end{eqnarray*}
Using a standard diagrammatic approach,\cite{Cirano} one can derive 
a set of two coupled dynamic equations for 
$C_{ij}(t,t_w)$ and $R_{ij}(t,t_w)$ for {\it any} model. In 
general though,  one does not suceed in performing the sum of diagrams 
involved.
There are two main strategies to obtain a closed set of equations.
The first consists in using simple models, i.e. simple expressions for 
$E({\vec \phi})$. This is the choice made when working with 
$O(N)$-type models for ferromagnetism,\cite{Bray} fully-connected 
spin-glass models,\cite{Spin-glassstatics,review,Gus}
models in infinite dimensional embedding spaces,\cite{manifold,Kotliar} etc. 
The second consists in choosing a recipe to select a subset of the infinite
set of diagrams in such a way that one can sum them all and express
this sum as an explicit functional of $C_{ij}$ and $R_{ij}$. This is a route commonly 
followed in field theory and it is used, for example,  to derive the mode-coupling theory for 
super-cooled liquids.\cite{Gotze,review}  
    
In fully-connected models all $n$-point functions can be expressed in terms of two-point 
functions, which then provide  a complete description
of the dynamics. In finite dimensional models this cannot be done in  closed, exact  form,  
and the description in terms of two-point functions is not complete.  

For all models,
the equations for two-time functions have  the structure of Schwinger-Dyson 
equations.  Our  claim is that in the SEP limit  their solutions share the aspects
discussed in Sect.~\ref{dynamic-scenario}. This is a consequence of the spontaneous 
breaking of a large reparametrization invariance appearing in the long waiting-time 
limit.\cite{Cuku-stat}

We wish to stress that neither the methods used to 
obtain the dynamic equations nor the scenario discussed in this 
paper rely on the presence of quenched disorder.  
This is unnecessary in a dynamic treatment.  

\section{The dynamic scenario}
\label{dynamic-scenario}

\subsection{Fast and slow dynamics. Infinite sensitivity of the slow dynamics}

Two-time functions have different behaviours in two
time-regimes separated by a model-dependent characteristic time 
${\cal T}(t_w)$ that diverges in the SEP limit.\cite{Cuku} 
The relaxation is fast for $t-t_w$  
small with respect to ${\cal T}(t_w)$ and it 
is fast for $t-t_w$ large with 
respect to ${\cal T}(t_w)$, $t-t_w\geq 0$. 
The correlation
and response are then
\begin{eqnarray*}
  C_{ij}(t,t_w) &=& 
C_{ij}^F(t-t_w) + {\tilde C_{ij}}(t,t_w)
\; ,
\nonumber\\
R_{ij}(t,t_w) &=& R_{ij}^F(t-t_w) 
 + {\tilde R_{ij}}(t,t_w)
\; ,
\end{eqnarray*}
where $C_{ij}^F$ and $ R_{ij}^F$ are nonnegligible only for small time-differences, 
and ${\tilde C_{ij}}$ and ${\tilde R_{ij}}$ vary very slowly.
This separation becomes sharper and sharper in the SEP limit. In this sense our
treatment is asymptotic.

A very general feature of systems with slow dynamics is that
{\it the time-dependence} of ${\tilde C_{ij}}$ and ${\tilde R_{ij}}$ 
{\it is sensitive
to vanishingly small changes in the equations of motion}. The most familiar example is
ferromagnetic coarsening in which an arbitrary small random field changes the 
growth law\cite{Na}
from power to log, 
making the slow part of the autocorrelation change from being a function of $t_w/t$ to
being a function of $\ln(t_w/\tau_0)/\ln(t/\tau_0)$. 
Another extreme example is the case of mean-field glasses,  in which
weak  nonconservative forces completely destroy aging, rendering the problem 
stationary.\cite{Cukulepe}

The ``infinite sensitivity'' is due to the presence of flat 
directions in the free-energy  landscape\cite{Cuku,Kula,Biroli} and
this is also the  very reason why there is  slow dynamics.
This physical fact has a mathematical counterpart in the 
invariances of the equations.\cite{Cuku-stat} 

The fact that one can change from aging to stationary slow dynamics
by applying infinitesimal perturbations is indeed the justification for treating both situations
on the same footing.

\subsection{Fluctuation-dissipation theorem 
and natural effective temperatures: time-scales and thermalisation.}

Quite generally,
a weak form of the  fluctuation-dissipation theorem (FDT) holds
in the SEP limit\cite{Cudeku}
\begin{equation}
R_{ij}(t,t_w) -\beta  \frac{\partial C_{ij}(t,t_w)}{\partial t_w} \; 
\Theta(t-t_w) \sim 0
\; .
\label{cosa22}
\end{equation}
This is either because both terms are large and cancel leading to the usual FDT
or because both are small but not necessarily equal, violating FDT. 
$\beta$ is the inverse temperature of the thermal bath, itself in equilibrium.
A stronger relation holds also in the SEP limit:
\begin{equation}
R_{ij}(t,t_w)= \beta_{ij}^{\sc eff} (t,t_w) \frac{\partial C_{ij}(t,t_w)}{\partial t_w} \; \Theta(t-t_w) 
\label{defi}
\end{equation}
and defines the inverse `effective temperature'\cite{Hosh,footnote3,Cukupe,Sciortino}
$\beta_{ij}^{\sc eff}$. Equation (\ref{cosa22}) states that
one can  have effective temperatures that are different  from $1/\beta$
only for the slow degrees of freedom. In what follows we call {\it natural effective 
temperatures}  
the values that $T^{\sc eff}$ takes for a system in contact with a {\it single}
bath, as opposed to those 
arising due to more complicated thermal environments.
The existence of non-trivial effective temperatures 
is essentially related to the production of entropy.\cite{Cudeku} 

If we define, as in experiments, \cite{aging_spinglasses}
\hspace{-1cm}
\begin{eqnarray}
(\chi_{ij}'+i\chi_{ij}'')(\omega,t_w) &=&
 \int_0^{t_w} d\tau e^{i \omega \tau} R_{ij}(t_w,t_w-\tau) 
\label{chidef}
\\
\tilde C_{ij}(\omega,t_w) &=& \int_0^{t_w} d\tau e^{i \omega \tau} C_{ij}(t_w,t_w-\tau)
\label{ctildedef}
\end{eqnarray}
$\beta^{\sc eff}$ can be alternatively defined by using\cite{Cukupe}
\begin{equation}
\beta^{\sc eff}_{ij}(\omega,t_w)
\equiv
\frac{\chi_{ij}''(\omega,t_w)}{\omega {\mbox{Re}} \tilde C_{ij}(\omega,t_w)}
\; .
\label{defi1}
\end{equation}
Equations (\ref{defi}) and (\ref{defi1}) do not necessarily 
coincide.

The scenario discussed  in this paper states that 
{\em for all pairs of observables and 
for each time-scale the effective temperatures
take a single value}.
That is,
\begin{eqnarray}
\beta_{ij}^{\sc eff}(\omega,t_w) 
&\sim& \beta^{\sc eff}(\omega,t_w) \; ,\forall \; i,j
\label{terma}
\\
\beta^{\sc eff}(\omega_1,t_w) &-&
\beta^{\sc eff}(\omega_2,t_w) \sim 0 
\;, {\mbox{if}} \; 
\frac{\omega_1}{\omega_2}={\mbox{cst}}
\label{scala}
\end{eqnarray}
in the SEP (old aging or weakly driven) limit.
Importantly enough, the definitions (\ref{defi}) and (\ref{defi1}) 
are in this limit  completely equivalent.
[Note, in passing,  a peculiarity of the (important) case of infinite effective
temperatures: all $\beta$'s in (\ref{terma}) and (\ref{scala})  
might tend to zero but their quotient not be one.]

It is also important to mention that the value of $T^{\sc eff}$ can sometimes 
very slowly change in time, in a much slower scale than all other in the problem, as emphasized 
by Nieuwenhuizen.\cite{Ni}

\begin{figure}
\centerline{\hbox{
\epsfig{figure=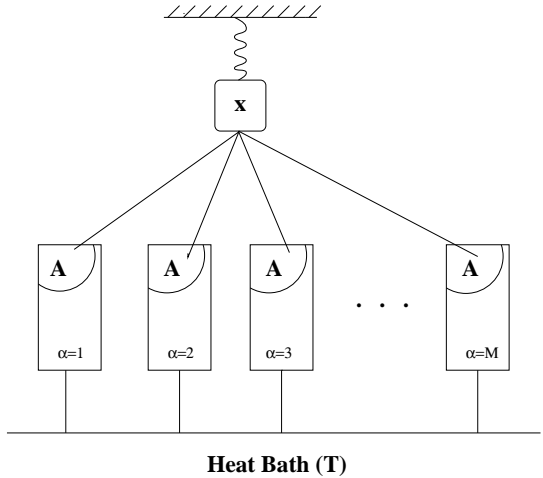,width=8cm}}
}
\vspace{-0.5cm}
\caption{$M$ copies of the system coupled to a heat bath at temperature $T$
and to a thermometer that is, in this sketch, 
represented by an oscillator.
}
\label{thermometer.fig}
\end{figure}

In order to show that $\beta^{\sc eff}(\omega_1,t_w)$ 
is indeed a temperature
measured by a thermometer we proceed as in 
Ref.~\cite{Cukupe,footnote5}. 
We  measure, with a `thermometer'
consisting of a degree of freedom $x$ with  a potential 
$V(x)$ (see Fig.~\ref{thermometer.fig}), 
 the temperature of an observable $A({\vec \phi})$.
In order to do so, 
we couple the thermometer linearly to the observable
 $A$ of an ensemble of many ($M$) independent copies
of the system, subject to the same history.\cite{Barrat}
  In App.~\ref{Athermometer.app} we show that the equation of
motion for  $x$ becomes, for large $M$,
\begin{equation}
m {\ddot x}(t) = -\frac{\partial V(x)}{\partial x(t)} + k^2 \int_{t_0}^t ds \, 
R_{AA}(t,s) x(s) + \rho(t)
\label{lange1}
\end{equation}
with $\rho$ a Gaussian noise with zero mean and correlation
\begin{equation}
\langle \rho(t) \rho(t_w) \rangle = k^2 C_{AA}(t,t_w)
\; .
\label{lange2}
\end{equation} 
$C_{AA}$ is the auto-correlation of the observable $A$ at two different times and 
$R_{AA}$ is the response of the same observable with respect to an infinitesimal  
perturbation $h$ 
that acts at time $t_w$ modifying the Hamiltonian according to 
$H \rightarrow H - A(t_w) h$. 
If $R_{AA}$ and $C_{AA}$ are linked by a time-scale dependent  
effective temperature that satisfies 
(\ref{scala}), we can decompose them 
as a sum of terms, each evolving in  a different  time-scale.
Hence, in the weak coupling limit $k^2\to 0$, 
Eq.~(\ref{lange2})  is 
a Langevin equation with several `baths' of the form (\ref{baba})
acting on widely separated time-scales and 
each with its own temperature. 
If $x$ is tuned to respond to a {\it single} time-scale,
it will measure only its temperature.\cite{exartier}

A simple way to view the relation of $\beta^{\sc eff}$
to the zeroth law\cite{Cukupe}  is
to  connect alternatively $x$ to an observable $A$ and an 
observable $B$ (simultaneously
for an ensemble of $M$ systems, as in the thermometer case).
One shows that the average heat flow goes
from the observable having a lower to the one having a 
higher $\beta^{\sc eff}$.

On the experimental side,
Grigera and Israeloff\cite{Tomas} gave evidence for the 
existence of non-trivial effective temperatures in 
glycerol by showing that Nyquist FDT 
is violated below $T_g$. In the context of 
spin-glasses, an indirect  -- and not exact -- method also suggests that non-trivial 
effective temperatures appear in the glassy phase.\cite{Cugrkuvi}

On the numerical side, the simulations of many groups gave evidence for the 
existence of non-trivial FDT violations (and, consequently, effective 
temperatures) in a large variaty of models. Among them we can 
mention models undergoing domain growth,\cite{domain-growth}
spin-glasses in finite dimensions,\cite{spin-glasses} and  
Lennard-Jones systems.\cite{lenard-jones} 

\subsection{Response to slow, auxiliary baths. Thermalization of subsystems.}

If glassy systems have natural effective temperatures 
associated with the slow degrees of freedom,
it is then natural to ask how would  
they react to a small coupling to an additional slow 
auxiliary bath of temperature $T^*$.\cite{Cuku-stat} 
The effect of the additional bath is taken into 
account by adding a term like (\ref{baba}) to the Langevin equation. 
The simplest choice is $\nu(t,t_w)={\tilde \nu}(|t-t_w|/\tau^*)$ with 
$\tau^*$ the characteristic time of the bath and ${\tilde \nu}$ 
a smooth decreasing function. 

Within this scenario an arbitrarily weak  auxiliary 
bath with  ${\tilde \nu}(0)$ small has an important effect 
provided it is sufficiently slow ($\tau^*$ large), and the 
temperature $T^*$ is within the range of effective temperatures 
of the slow dynamics. 
The (slow) time dependences of all correlation and response 
functions are affected by  a time rescaling $t \rightarrow K(t)$
such that the time-scale which has a natural effective 
temperature $T^*$ is locked to the time-scale $\tau^*$.
(This is possible thanks to the sensitivity property.) 

In particular, an aging system with a single $T^{\sc eff}>T$ will become 
stationary if $T^*>T^{\sc eff}$, and be hardly affected 
if $T^*<T^{\sc eff}$. More generally, 
an aging  system with multiple effective temperatures 
will  become partially stationary (for  
all the time-scales with $T^*>T^{\sc eff}$), 
but will still have aging for time-scales with 
$T^*<T^{\sc eff}$. 

We  hence have argued,\cite{Cuku-stat} very much in the spirit of 
Refs.~\cite{virasoro,pinning,Remi,Allah} that
the coupling to slow auxiliary baths are 
perturbations `conjugate to the natural temperatures'.
The arguments put forward in the previous two paragraphs are supported by 
simulations of spin-glasses
 both mean-field and 3D (the latter very encouraging).\cite{Cuku-stat}
In App.~\ref{app.Monte} we describe the algorithm used to simulate
a spin system in contact with a multiple bath.

Another  relevant question is   
mutual thermalisation between two subsystems 
having different natural effective temperatures, but sharing the same bath.\cite{Cukupe} 
This is related to the response to an auxiliary slow  bath,
since, to a certain extent,
each system acts as an auxiliary bath for the 
other when placed in mutual contact.

Within the present scenario   
two  situations are possible:\cite{Cuku-stat}  strong coupling, in which
the effective temperatures equalise; and weak coupling, 
in which the combined system 
preserves essentially the temperatures  of its constituents, 
but rearranges them 
in widely separated time-scales. In the latter case, 
the combined system has three temperatures, 
the bath  temperature for the fast processes, the lowest  and highest
original
effective temperatures for  the intermediate and slowest timescales, 
respectively.

\subsection{Reciprocity relations.}

An unexpected feature that appears in this scenario are 
the reciprocity (Onsager) relations:\cite{Cuku-stat}
\begin{equation}
\langle A(t) B(t_w) \rangle = 
\langle B(t) A(t_w) \rangle
\label{reciprocity}
\; .
\end{equation}
If this equality holds for the observables $A$ and $B$, 
Eqs.~(\ref{terma}) and (\ref{scala}) imply a similar relation
for the responses.
There is no reason {\em a priori} why the reciprocity relations 
should hold in a situation in which FDT is strongly violated. 
This is all the more surprising in an
aging case in which the system is not even stationary. 
The interest of these relations is that they are relatively easy to measure
in a simulation (results for 3DEA appeared in Ref.~\cite{Cuku-stat}) or in an experiment. 
The reciprocity relations are intimately related to the partial equilibrations. 
.

\section{Multiply thermalized systems}
\label{multiple.sect}

A standard procedure in the dynamics of extended systems is to keep some 
macroscopic variables of interest and integrate away
the others. The eliminated variables become 
part of a `thermal' bath.

In {\it equilibrium}, the validity of FDT 
and of time-translational invariance (TTI) ensure that
the friction and noise terms, coming from the projected sector, 
satisfy FDT (of the second kind).
The whole procedure is self-consistent since the 
variables evolving in contact with a bath
satisfying FDT will eventually satisfy FDT, now called of 
the first kind.\cite{Kubo}

In an {\it out of equilibrium} situation
FDT (and sometimes TTI) do not hold.
How can we then picture the effect of 
the projected variables on our chosen ones?

We treat a system in contact with a multiple
bath with different temperatures and time-scales that is based on subsequent 
adiabatic approximations of the slower baths.
This approach has points in common with the 
`pinning field' approach of Kirkpatrick and Thirumalai\cite{pinning} and
Monasson\cite{Remi} and with recent developments of 
Allahverdyan {\it et al.}\cite{Allah} and Franz and Virasoro\cite{silvio}
We claim that, 
in the SEP limit, 
{\em correlation and response functions of the selected macroscopic variables
behave as if they were subject to such a multiple bath}.

By studying in detail the evolution of a harmonic oscillator  
in contact with a multiple bath,
we show how, even in this very simple model, different effective temperatures
are induced by the coupling to several baths.\cite{exartier2}

\subsection{The construction}
\label{Theconstruction.subsection}

Consider a particle of coordinate $x$ moving in a potential $V(x)$ under
the influence of two thermal baths:
a fast white bath  of temperature $T$ 
and friction coefficient $\Gamma_0$,
and a slow bath of temperature $T^*$, characteristic time-scale $\tau^*$
and strength $\Gamma_1$.
Note that such a system has a non-Gibbsean stationary  measure.
We are interested in the evolution of the particle taking  
$\tau^*$ large with respect to any other characteristic time.
(The following argument can be easily extended to any number 
of variables $x_i$, $i=1,\dots,N$.)
This problem, and its solution, is very similar to the one studied By Allahverdyan and Nieuwenhuizen 
 in the second article of \cite{Allah}. These authors analysed  two interacting variables that 
are respectively coupled to two  baths with different temperatures and very 
different time-scales.

Neglecting the effect of inertia one has
\begin{eqnarray*}
\Gamma_0 {\dot x}(t) + \int_{-\infty}^{t}  ds \; \tilde \nu(t-s) {\dot x}(s)
=
-\frac{\partial V}{\partial x(t)} + \eta(t) + \rho(t)
\; ,
\label{uno1} 
\end{eqnarray*}
where $\eta$ and $\rho$ are the Gaussian thermal noises of the fast and
slow baths, respectively, both with zero mean and variances
$\langle \eta(t) \eta(t_w) \rangle = 2 T \Gamma_0 \delta(t-t_w)$
 and
$\langle \rho(t) \rho(t_w) \rangle=T^*\tilde\nu(|t-t_w|/\tau^*)$.
We assume that
$\tilde \nu(0) \equiv \Gamma_1 
$ and 
$ {\dot{\tilde\nu}}(0^+)=0$.
We have set the system in contact with the slow bath at the 
initial time $t_0=-\infty$.
If we integrate by parts, 
\begin{eqnarray}
\Gamma_0 {\dot x}(t)  &=&
-\frac{\partial V}{\partial x(t)} - \Gamma_1 x(t) + \eta(t) + h(t) 
\label{dos}
\\
h(t) &\equiv& - \int_{-\infty}^{t}  ds \; {\dot{\tilde\nu}}(t-s)  x(s)  + \rho(t)
\; .
\label{tres} 
\end{eqnarray}
In the adiabatic limit, 
the slow bath generates a quasi-static 
field $h(t)$. (Indeed, using ${\dot{\nu}}(0^+)=0$ and $\tau^* \gg 1$, 
one shows ${\dot h}=O(1/\tau^*) \ll 1$.) 
Hence, $x$ has a fast evolution 
given by Eq.~(\ref{dos}) with $h$ fixed and it
achieves a distribution
\begin{equation}
P(x/h)= \frac{ 
e^{-\beta \left( V(x)+\Gamma_1 \frac{x^2}{2} - hx \right)}
}
{
\int \; dx \; 
e^{-\beta \left( V(x)+\Gamma_1 \frac{x^2}{2} - hx \right) }
}
\; .
\label{ll}
\end{equation}
The denominator defines $Z(h)$ and 
$F(h) \equiv -\beta^{-1} \ln Z(h)$.
Henceforth we denote $P(a/b)$ the conditional
probability of $a$ given $b$ at stationarity.

The approximate evolution of $h$ 
is given by Eq.~(\ref{tres}) with the replacement of 
$x$ in the friction term by its average with respect to the fast evolution:
\begin{equation}
h(t) = \int_{-\infty}^{t} ds \, {\dot \nu}(t-s)
  \frac{\partial F(h)}{\partial h}(s)  + \rho(t)
\; .
\label{cinco} 
\end{equation}
This equation is a non-Markovian and it has all the properties of a system coupled to a (slow) bath
of temperature $T^*$. In particular, the stationary distribution is
\begin{equation}
{\hat P}(h)=\frac{ e^{-\beta^* (F(h)+ \frac{h^2}{2\Gamma_1})} }
{\int \; dh \; e^{-\beta^* (F(h)+ \frac{h^2}{2\Gamma_1})}}
\; .
\label{gg}
\end{equation}
Reciprocity and FDT for functions of $h$ also hold 
at stationarity.
The stationary distribution of $x$ now reads
\begin{equation}
P(x)= \int  dh \, P(x/h){\hat P}(h)   
\; .
\label{conditional}
\end{equation}
We prove these properties in App.~\ref{pinning.app}.
In particular, if $T=T^*$ 
we recover the usual GB distribution
for $P(x)$ regardless of the details of the slow bath.

Note that the distribution function $P(x)$ contains the superposition 
of different time-scales and is not a very eloquent 
quantity.

 A case of great  interest is the one of a 
slow bath that is not itself stationary, but ages. 
Suppose that we have
\begin{equation}
\nu(t,s) = \tilde\nu\left(L(s)/L(t)\right) \;\;\;\;\;\;\;\;\; t \geq s
\end{equation}
for some monotonically increasing function $L$.
For the fast motion, at large enough times,
we recover Eq.~(\ref{ll}) while
Eq.~(\ref{tres}) follows from the change of variables 
\begin{equation}
{\cal T}= \ln{L(t)} \; , \;\;\;\;\;\;\; {\cal T}'= \ln{L(s)}
\; .
\label{tauu}
\end{equation}
In these new time-like variables, the rest of the derivation can be carried through
identically.
`Stationary' means in this case invariant with respect to $L
\rightarrow \Delta \times  L$
that leads to ${\cal T} \rightarrow {\cal T} +\ln \Delta$ in time-like 
variables ($\Delta$ is a parameter). 

\subsection{Reciprocity and FDT}
\label{reciprocityandFDT}

In its strongest form, reciprocity means that the joint probability of
having $x$ at time $t$ and $x'$ at $t_w$ is equal 
{\em at stationarity}\cite{footnote1} to 
the probability of
having $x'$ at time $t$ and $x$ at $t_w$. To prove that this indeed happens, we use
the  separation of time-scales and
consider two cases:

{\it i.}
If $t-t_w$ is small, $h$ is approximately fixed and we have an 
equilibrium problem at temperature $T$, with a fixed field $h$. Reciprocity holds in
the usual way.

{\it ii.}  If $t-t_w$ is such that $h(t)-h(t_w) \neq 0$, we have
\begin{eqnarray*}
P(x,t;x',t_w)&=& \int dh dh' 
P(x/h)  P(h,t / h',t_w) P(h'/x')  P(x')
\label{rec}
\end{eqnarray*}
Since $
 P(h, t ; h',t_w) =  P(h',t; h,t_w)
$ (see App.~\ref{pinning.app}), 
\begin{equation}
P(x,t;x',t_w)=P(x',t;x,t_w)
\; .
\label{recipro}
\end{equation}

In a similar way one shows that FDT holds separately for each
timescale with its own temperature.

{\it i.}
If $t-t_w$ is small FDT holds
for each $h$. Denoting $\delta P(x,t)/\delta H_A(t_w)$ the variation of the distribution
at time $t$ due to a field conjugate to $A$ that has been on from $-\infty$ to $t_w$, we have
\begin{equation}
\frac{\delta P(x,t)}{\delta H_A(t_w)} = \beta \int dx' \, P(x,t/x',t_w) A(x') P(x') 
\; .
\end{equation}
The integrated form of FDT 
for two observables $A$ and $B$ can be obtained 
by multiplying by $B(x)$ and integrating over $x$.

{\it ii.}  If $t-t_w$ is such that $h(t)-h(t_w) \neq 0$,
while the field conjugate to $A$ is on, $h$ evolves with Eq.~(\ref{cinco}) 
and
\begin{eqnarray}
F_A(h) &=& - \frac{1}{\beta} 
\ln \int dx' e^{-\beta \left( V(x') -H_{A} A(x') +
\Gamma_1 \frac{{x'}^2}{2} - hx' \right)}
\nonumber \\
&=&  F(h) -  H_A {\cal A}(h)
\; , 
\end{eqnarray}
with ${\cal A}(h)=\int dx'  P(x'/h) A(x') $.
If $H_A$
is turned off at $t_w$, $P(x/h)$,  
is first modified in a short time-scale
with respect to that of $h$. Then, $h$ continues to evolve but with a modified
effective potential $F(h)$.
Now, FDT holds  for the evolution of $h$ at temperature $T^*$ 
\begin{eqnarray*}
\frac{\delta {\hat P}(h,t)}{\delta H_A(t_w)} = 
\beta^* \int dh' \, P(h,t/h',t_w) {\cal A}(h') {\hat P}(h') 
\; ,
\end{eqnarray*}
(see App.~\ref{pinning.app}) and 
\begin{eqnarray*}
& & \frac{\delta P(x,t)}{\delta H_A(t_w)} = \int dh \; 
P(x/h) \frac{\delta {\hat P}(h,t)}{\delta H_A(t_w)}  \nonumber \\
 &&=  \beta^* \int dh \; dh'\; P(x/h) P(h,t/h',t_w) {\cal A}(h') {\hat P}(h') 
\; .
\end{eqnarray*}
Using ${\hat P}(h') {\cal A}(h')  
= \int dx' P(x') P(h'/x')A(x')
$
we obtain
\begin{eqnarray*}
\frac{\delta P(x,t)}{\delta H_A(t_w)} &=&
\beta^* \int  dx' \, P(x,t/x',t_w) A(x') P(x')
\; .
\end{eqnarray*}
For separate times FDT holds with temperature $T^*$.

\subsection{The effect of a weak coloured bath}

The simplest example where to check the ideas described in 
Sect.~\ref{Theconstruction.subsection} and  \ref{reciprocityandFDT} 
is a harmonic oscillator in contact with a white and a 
coloured bath with exponential correlation. This problem can be tackled 
using a variety of techniques.

\begin{figure}
\centerline{\hbox{
\epsfig{figure=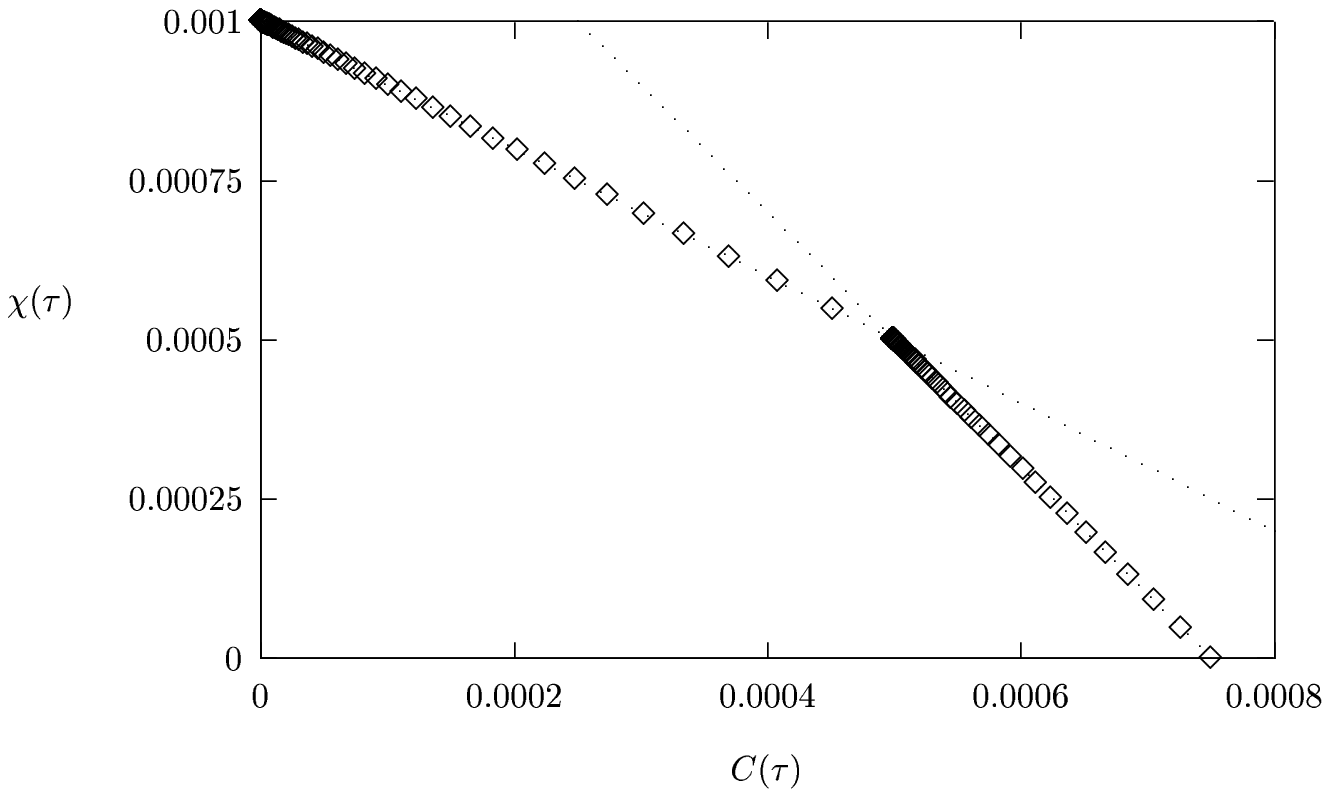,width=7cm}
}}
\vspace{-.5cm}
\centerline{\hbox{
\epsfig{figure=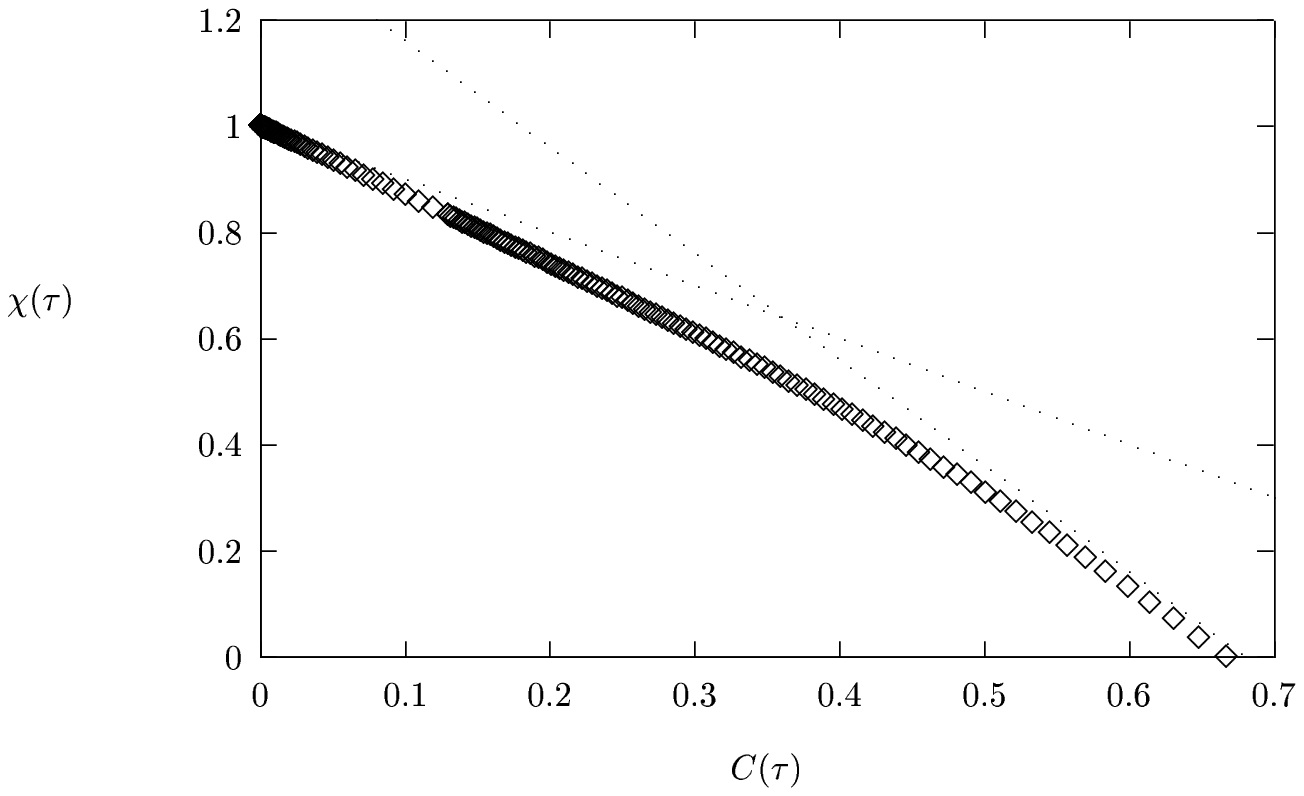,width=7cm}
}}
\vspace{-.5cm}
\caption{The $\chi(C)$ plot for a harmonic oscillator 
subject to a white bath of temperature $T=0.5$ and a coloured
noise of temperature $T^*=1$. Above: $(\omega+\Gamma_1)\tau^*=2000 \gg \Gamma_0=1$.  
The straight lines (dots) have inverse slopes 
$-1/T=-2$ and $-1/T^*=-1$. Below:
$(\omega+\Gamma_1)\tau^*=2=O(\Gamma_0=1)$. 
}
\label{osc}
\end{figure}

From the exact asymptotic solution
we extract the behaviour of the integrated response
$\chi(\tau) \equiv \int_0^{\tau} d\tau' \, R(\tau')$ vs. 
$C(\tau)$ and read $T^{\sc eff}$ from it.
In Fig.~\ref{osc} we show $\chi(C)$
for $T=0.5$, $\Gamma_0=1$, $T^*=1$.
In the plot above, $(\omega+\Gamma_1) \tau^* =2000 \gg \Gamma_0=1$, 
with $\omega$ the frequency of the oscillator. 
The evolution takes 
place in two time-scales (or correlation-scales) characterised by temperatures 
$T$ and $T^*$.
The straight lines have slopes $-1/T$ and $-1/T^*$
showing that there are two temperatures associated
to the motion of the particle: a fast motion for $q< C<q_d$  controlled by $T$ 
and a slow motion for $0< C< q$ that is instead controlled by $T^*$. 
In the plot below $(\omega+\Gamma_1) \tau^* = 2=O(\Gamma_0=1)$, 
the time-scales (or correlation-scales) are not well separated,  
and $\chi(C)$ 
continuously interpolates between a region of slope of $-1/T$  
and a region of slope of $-1/T^*$.

This problem can be alternatively  analysed using the technique described in 
Sect.~\ref{Theconstruction.subsection}. 
One recovers, as expected, the 
exact results of the previous paragraphs. 

\subsection{Many scales}

The construction of Sect.~\ref{Theconstruction.subsection} can 
be generalised to many nested scales. 
Let us see how this is done for three baths;
the equation of motion is
\begin{eqnarray}
\Gamma_0 {\dot x}(t) &+& \int_{-\infty}^{t} ds 
\left( \nu_1(t-s)+ \nu_2(t-s) \right) {\dot x}(s)  =
-\frac{\partial V(x)}{\partial x(t)} 
\nonumber\\
& & + \eta(t) + \rho_1(t) + \rho_2(t)
\label{unos} 
\end{eqnarray}
with $\rho_1$ and $\rho_2$ two Gaussian thermal noises with 
zero mean, 
$\langle \rho_1(t) \rho_1(t_w) \rangle = T^*_1 {\tilde \nu}_1(|t-t_w|/\tau^*_1) $, 
$\langle \rho_2(t) \rho_2(t_w) \rangle = T^*_2 {\tilde \nu}_2(|t-t_w|/\tau^*_2)$
and $1\ll \tau^*_1 \ll \tau^*_2$.
Proceeding as before, we write
\begin{eqnarray*}
\Gamma_0 {\dot x}(t)  =
-\frac{\partial V(x)}{\partial x(t)} + (\Gamma_1+\Gamma_2)\,  x(t) + \eta(t) + h_1(t) + h_2(t)
\end{eqnarray*}
with
$
h_i(t) \equiv - \int_{-\infty}^{t}   ds \; {\dot \nu}_i(t-s)  x(s)  + \rho_i(t) 
$, $i=1,2$,
and
$
\nu_i(0) = \Gamma_i$, 
$\dot \nu_i(0^+) = 0$,
$i=1,2$.

In the fastest evolution  $x$ achieves 
distribution,
\begin{equation}
P(x/h_1,h_2)= \frac{ \, 
e^{-\beta \left( V(x)+( \Gamma_1+\Gamma_2)  \frac{x^2}{2} - (h_1+h_2)x \right)}
}
{Z_{h_1,h_2}}
\; ,
\label{lls}
\end{equation}
with $Z_{h_1,h_2}$ ensuring $\int dx \, P(x/h_1,h_2)=1$.
The  evolution of $h_1$ is obtained by taking $h_2$ to be adiabatic
\begin{equation}
h_1(t) = \int_{-\infty}^{t}  ds \; {\dot \nu}_1(t-s)
  \frac{\partial F_1(h_1+h_2)}{\partial h_1}(s) + \rho_1(t)
\label{cincos1} 
\end{equation}
with
$
e^{-\beta^*_1 F_1(h_1+h_2)} \equiv 
\int dx \; 
e^{-\beta \left( V(x)+ (\Gamma_1+\Gamma_2) \frac{x^2}{2} - (h_1+h_2)x \right)}
$
yielding a stationary distribution for $h_1$
\begin{equation}
P(h_1/h_2) \propto 
\exp\left[-\beta^*_1 \left( F_1(h_1+h_2)+ \frac{h_1^2}{2 \Gamma_1^2} \right)\right]
\; .
\label{llss2}
\end{equation}
Finally, $h_2$ evolves at inverse temperature $\beta^*_2$ and has a stationary distribution:
\begin{equation}
{\hat P}(h_2)\propto 
\exp\left[ -\beta^*_2 \left(F_2(h_2)+  \frac{h_2^2}{2 \Gamma_2^2} \right) \right] 
\end{equation}
with
$
e^{ -\beta^*_2 F_2(h_2)} \equiv 
\int dh_1 \; 
e^{-\beta_1^* \left(      F_1(h_1+h_2)+ \frac{h_1^2}{2 \Gamma_1^2}
                          \right)}
$.
 One can also consider
two slow baths that are non-stationary. 
In order to have two well separated time-scales, 
we need two (increasing) scaling functions $L_1(t)$,  $L_2(t)$
with 
\begin{equation}
\lim_{t \rightarrow \infty}
\left( 
d_t \ln L_2(t)/d_t \ln L_1(t) 
\right)
= 0 
\end{equation}
and, for example, 
\begin{eqnarray*}
\nu_1(t,s) = \tilde \nu\left(L_1(s)/L_1(t)\right) \; ,
\nu_2(t,s) = \tilde \nu\left(L_2(s)/L_2(t)\right) \; , 
\label{twoscales}
\end{eqnarray*}
$t \geq s$. After performing two transformations like the ones in Eq.~(\ref{tauu}),
the rest of the derivation can be repeated.

For this example one concludes that
the `probability cloud' of each scale 
acts as a configuration for  the slower scale --- moving in an 
effective potential.

\section{Conclusions}
\label{conclusions.section}

In the last years great progress has been made in understanding the 
asymptotic nonequilibrium evolution of systems with slow dynamics.
Much of these developments came from the study of simplified models
that include classical glassy disordered and non-disordered models, 
quantum disordered systems
and, importantly enough, systems that are constantly driven out of
equilibrium by external forces. 
The results collected from the solution of several of these cases 
made apparent the existence of various general features. 
In this article (as well as in Ref.\cite{Cuku-stat})
we stressed which are the features that we believe
build a scenario for the slow dynamics
in the limit of small entropy production.
Properties {\it i}-{\it iv} discussed in Sect.~\ref{dynamic-scenario}. 
 can be quite surprising, in particular the latter.
In addition, in view of the infinite sensitivity of the slow time dependence 
of the correlation and response functions, it is a non-trivial fact 
that properties {\it ii}-{\it iv} are preserved under small perturbations.  
In Ref.~\cite{Cuku-stat} we justified this fact by studying the invariances
(and breakdown of) the dynamic equations in the SEP limit.\cite{footnote6} 

In short, the image we have in mind is better grasped by comparing the 
general case to the simple problem studied in Sect.~\ref{multiple.sect}.  
The Langevin equation for a variable $x$ including 
a non-local friction kernel and a correlated noise, Eq.~(\ref{uno1}), is
reminiscent of the equation for a single effective variable
that one obtains in fully connected models, large dimension approximations, 
etc.\cite{Kotliar,Sozi} 
Indeed, in the large $N$ limit, one can 
derive such an equation by performing a saddle-point approximation 
of the dynamic generating functional. 
For any dynamic variable $\phi_i$, the `single-variable equation' 
reads
\begin{equation}
\Gamma_0 \, \dot \phi_i(t) = -
\mu(t) \,  \phi_i(t) + \int_{t_0}^t ds \, \Sigma(t,s) \; \phi_i (s) + \rho_i(t)
+\eta_i(t)
\; .
 \label{9}
\end{equation}

There are two noise sources in this equation: $\eta_i(t)$ is the 
original white noise while $\rho_i(t)$ is an effective (Gaussian) noise with zero mean and 
correlations self-consistently given by $\langle \rho_i(t) \rho_j(t_w) \rangle = \delta_{ij}  T^*
D(t,t_w)$. 
The vertex $D(t,t_w)$ plays the r\^ole 
of the coloured noise correlation in a usual Langevin 
equation.
The self-energy $\Sigma(t,t_w)$ appears here as an 
`integrated friction'.
 
In the case of a system in contact with an external coloured bath, 
the correlation of the noise $T^* \nu(t,t_w)$ is simply related to the retarded 
friction $\nu(t,t_w)$. 
This is the statement of 
FDT applies for the bath.
In the case of the single-variable  equation for a more complicated model, 
the  friction kernel $\Sigma(t,t_w)$ and the correlation of the coloured 
noise $T^* D(t,t_w)$ are properties of the system and are not
necessarily related in a simple manner. In the SEP limit, 
they get related by a modification of FDT with time-scale dependent
effective temperature $T^{\sc eff}(t,t_w)$. 

One then concludes that the 
structure of these two problems is indeed very similar:
\begin{itemize}
\item
If one weakly couples a simple system to a slow bath of temperature $T^*$, 
at sufficiently slow time scales the system acquires the temperature $T^*$.
\item 
Glassy systems arrange their internal degrees of freedom in such 
a way that slow degrees of freedom select their own effective temperatures. 
\end{itemize} 

Implicit in the construction presented in Section~\ref{Theconstruction.subsection}
is the fact that all observables
have the same effective temperatures at the same two-times.
Also implicit are the reciprocity relations. These are important in that
they are potentially measurable, and quite unexpected out of equilibrium.

\appendix

\section{Measurements of natural effective temperatures}
\label{Athermometer.app}

We here recall the definition of  
a time-scale dependent effective temperature, for systems out of equilibrium, 
given in Ref. \cite{Cukupe}
The presentation follows closely this reference
but it makes some points of the derivation more precise.  

Let us consider 
$M$ non-interacting copies of the system,\cite{Barrat} and couple each of them  
to a simple system that acts as a thermometer in the manner sketched in 
Fig.~\ref{thermometer.fig}. For simplicity, 
we describe this thermometer with a single variable $x$ and 
each system with a variable ${\vec \phi}_\alpha$, $\alpha=1,\dots,M$.
The total energy of the complex is
\begin{eqnarray*}
E_{\sc tot}
= 
m \frac{{\dot x}^2}{2}+ V(x) 
+
\sum_{\alpha=1}^M E({\vec \phi}^\alpha) 
-
\frac{k}{M^{1/2}} x  
 \sum_{\alpha=1}^M A({\vec \phi}^{\alpha})
\; ,
\end{eqnarray*}
where $V(x)$ and $E({\vec \phi}_a)$ are the potential energies of  the isolated thermometer
and each isolated system.
For each copy the last term yields an infinitesimal 
(for $M$ large) perturbation 
corresponding to a field $k x/\sqrt{M}$ conjugate to the observable 
$A({\vec \phi}^{\alpha})$. The equation of
motion for $x$ is
\begin{equation}
m {\ddot x}(t) = -\frac{\partial V(x)}{\partial x(t)} - \frac{k}{M^{1/2}} 
\sum_{\alpha=1}^M A({\vec \phi}^{\alpha})(t)
\; .
\label{eqmotionx}
\end{equation}
For simplicity we choose an operator $A$ such that 
$\langle A({\vec \phi}^{\alpha}) \rangle_{k=0} =0$ where $\langle \bullet \rangle$ 
represents either an average over different histories of the same system 
or an average over different systems, e.g. 
$\langle f \rangle = 1/M \sum_{\alpha=1}^M f_\alpha$.
(If the average of the operator $A$ does not vanish, one has to work with 
the difference between the operator and its average.)
The subindex $k=0$ indicates that the average is taken in the absence of the 
external field $k x/M^{1/2}$. Whenever we take the average in the presence
of the field we shall denote it $\langle \bullet \rangle_k$. 

Assuming that linear response holds for each system,
the total variation of the average $\langle A({\vec \phi}^{\alpha}) \rangle_k$
caused by a field that has been applied from $t=0$ upto $t$ reads
\begin{eqnarray}
\delta \langle A({\vec \phi}^{\alpha}) \rangle_k(t) 
&=&
\langle A({\vec \phi}^{\alpha}) \rangle_k(t) -
\langle A({\vec \phi}^{\alpha}) \rangle_{k=0}(t) =
\nonumber\\
\langle A({\vec \phi}^{\alpha}) \rangle_k(t) 
&=&
\int_0^t ds \; R_{A_\alpha,A_\alpha}(t,s) \frac{k}{\sqrt{M}} x(s)
\; ,
\end{eqnarray}
where $R_{A_\alpha,A_\alpha}(t,s)$ is the linear response of the observable 
$A({\vec \phi}^\alpha)$
at time $t$ to a change in energy $-k/\sqrt{M} x A({\vec
  \phi}^\alpha)$ at time $s$. 
These thermal history-averaged  responses
are equal for all systems, 
we henceforth denote  them  $R(t,s)$ for all $\alpha$.  (We also simplify the 
notation by eliminating the subindex $A$ that identifies the observable.) 
Adding the last equation over $\alpha$ and multiplying by $k/\sqrt{M}$ we have
\begin{eqnarray}
\frac{k}{\sqrt{M}} \sum_{\alpha=1}^M 
\langle A({\vec \phi}^{\alpha}) \rangle_k(t) 
&=&
k^2 \int_0^t ds \, R(t,s) x(s)
\; .
\label{averageA3}
\end{eqnarray}
Adding and subtracting from the rhs of  Eq.~(\ref{eqmotionx}) 
the average (\ref{averageA3}), we recast it as
\begin{equation}
m {\ddot x}(t) = 
-\frac{\partial V(x)}{\partial x(t)} + k^2 \int_0^t 
ds \, R(t,s) x(s) + \rho(t)
\label{lange3}
\end{equation}
with 
\begin{equation}
\rho(t) \equiv 
\frac{k}{\sqrt{M}} 
\sum_{\alpha=1}^M  
\left( 
A({\vec \phi}^{\alpha})(t) - \langle A({\vec \phi}^{\alpha}) \rangle_k(t)
\right)
\; .
\end{equation}
The `force' $\rho(t)$ 
is  the sum of $M$ independent identically distributed random variables.
For large $M$  it becomes a Gaussian variable with zero mean and 
variance:
\begin{eqnarray}
\langle \rho(t) \rho(t_w) \rangle_k
&=& 
\frac{k^2}{M} \sum_{\alpha,\beta=1}^M  
\left. C_{\alpha\beta}^{\sc conn}(t,t_w)\right|_{k}
\; .
\end{eqnarray}
with $C_{\alpha\beta}^{\sc conn}(t,t_w)|_{k}=
\langle 
( A({\vec \phi}^{\alpha})(t) - \langle A({\vec \phi}^{\alpha}) \rangle_k(t) 
)$
$\times
( A({\vec \phi}^{\beta})(t_w) - \langle A({\vec \phi}^{\beta}) \rangle_k(t_w) 
)
\rangle_k
$
the connected correlation in the presence of the field.
Up to leading order in $k$ 
\begin{equation}
\left.
C_{\alpha\beta}^{\sc conn}(t,t_w)
\right|_{k}
 \sim 
\left.
C_{\alpha\beta}^{\sc conn}(t,t_w)
\right|_{k= 0}
\; .
\end{equation}
The connected correlation can also be substituted by the 
usual correlation since $A({\vec \phi}^{\alpha})(t)=O(1)$
implies $\langle A({\vec \phi}^{\alpha})(t) \rangle_k=O(1/\sqrt{M})$.
Thus
\begin{equation}
\left.
C_{\alpha\beta}^{\sc conn}(t,t_w)
\right|_{k=0}
 \sim 
\left.
C_{\alpha\beta}(t,t_w)\right|_{k=0}
\sim \delta_{\alpha\beta} C(t,t_w)
\end{equation}
since the bare systems are completely independent.
Finally,
\begin{equation}
\langle \rho(t) \rho(t_w) \rangle_k = 
k^2 C(t,t_w)
\; .
\end{equation}
Thus the dynamic equation governing the evolution of the thermometer reads
\begin{eqnarray}
m \ddot x(t) &=& -\frac{\partial V(x)}{\partial x(t)} + 
\rho(t) + k^2 \int_0^t ds \, R(t,s) x(s)
\label{eqqqq1}
\; ,
 \nonumber\\
\langle \rho(t) \rho(t_w) \rangle_k &=& k^2 C(t,t_w)
\; .
\end{eqnarray}
If the response and the correlation 
of the system are related through the FDT relation
\begin{equation}
R(t,s) = \frac{1}{T} \frac{\partial C(t,s)}{\partial s} \Theta(t-s)
\label{FDTsystem}
\; .
\end{equation}
one integrates by parts the last term on the rhs of Eq.~(\ref{eqqqq1})
and obtains
\begin{eqnarray}
m \ddot x(t) &=& -\frac{\partial V}{\partial x(t)} + \rho(t) - \frac{k^2}{T} \int_0^t ds \, C(t,s) \dot x(s)
\nonumber\\
& & 
+ \frac{k^2}{T} \left( C(t,t) x(t) - C(t,0) x(0) \right)
\; .
\label{eqqqq2}
\label{lange4}
\end{eqnarray}
that in the limit of weak coupling, $k^2 \to 0$, becomes a
Langevin equation since the last term
can be neglected.

From this equation we can conclude that the thermometer, being a system 
with fast dynamics,  will eventually 
equilibrate at the temperature that relates response and correlation of the system, 
Eq.~(\ref{FDTsystem}). 
{\it The system acts as a thermal bath on the thermometer.}

We can also derive this result by choosing a harmonic oscillator as a thermometer.\cite{Cukupe}
Using $V(x) = m \omega_o^2 x^2 /2$, 
one easily shows that the averaged potential energy of the thermometer over a time-window around the 
measuring time $t_w$ is
\begin{equation}
\frac{1}{2} \langle E_{\sc osc} \rangle_{t_w} 
=
\frac{1}{2} \omega^2_o \langle x^2 \rangle_{t_w} 
\; ,
\end{equation}
with $\omega_o$ is the 
probing frequency of the oscillator and $m=1$. 
In the large $t_w$ limit, it 
reaches the following limit
\begin{equation}
\langle E_{\sc osc} \rangle_{t_w}
= 
\frac{ \omega_o \tilde C(\omega_o, t_w) }{\chi''(\omega_o,t_w)} 
\end{equation}
with $\tilde C$ and $\chi''$ defined
in Eqs.~(\ref{chidef}) and (\ref{ctildedef}).
If equipartition holds for the 
oscillator, one has
\begin{equation}
T^{\sc eff}(\omega_o, t_w)
 = 
\frac{ \omega_o \tilde C(\omega_o, t_w) }{\chi''(\omega_o,t_w)} 
\; ,
\label{hola}
\end{equation}
i.e. the definition given in Eq.~(\ref{defi1}). 

Under the assumption (\ref{terma}) 
we can   decompose the correlations and responses 
as follows:
\begin{eqnarray*}
C(t,t_w)&=& C^F(t,t_w)+ C^1(t,t_w) + C^2(t,t_w)+ ... \nonumber \\
  R(t,t_w)&=& R^F(t,t_w)+ R^1(t,t_w) + R^2(t,t_w)+ ... 
\end{eqnarray*}
with
\begin{equation}
R^i(t,t_w)= 
\beta^i \frac{\partial C^i(t,t_w)}{\partial t_w}  \Theta(t-t_w) \;, 
\end{equation}
$i=0,1,\dots$, $i=0$ corresponding to the FDT scale.
Then, Eqs.~(\ref{lange2}) and (\ref{lange4}) 
corresponds to a system coupled
to a series of baths of inverse temperatures $\beta^F$, 
$\beta^1$, $\beta^2$ acting on 
widely separated scales. 
The probing frequency  of the 
thermometer can be selected to measure any of the values of the effective 
temperature.\cite{exartier}   

Actually, the effective temperature $T^{\sc eff}$ in Eq.~(\ref{hola}) 
might depend on the observable $A$ considered. The picture in this paper
describes the way in which we think these dependences arrange in a physical
system.

\section{Stationary distribution of the quasi-static field}
\label{pinning.app}

We show that Eq.~(\ref{cinco}) leads to Eq.~(\ref{gg})
and that at stationarity reciprocity and FDT hold.
The Fourier transform of Eq.~(\ref{cinco}) reads
\begin{eqnarray}
h(\omega)&=&r(\omega) \frac{\partial F}{\partial h} (\omega) + 
\rho(\omega) 
\; ,
\label{cincoff}
\\
\langle \rho(\omega) \rho(\omega') \rangle &=& 
T^* \nu(\omega) 2 \pi \delta (\omega + \omega') 
\; ,
\label{cincof}
\end{eqnarray}
where 
$
r(\omega)
 \equiv 
\int_{-\infty}^\infty d\tau \exp(i \omega\tau) 
\dot \nu(\tau) \theta(\tau)
$ is the slow bath's response function. It is then 
analytical in the upper half complex plane and satisfies,
for real $\omega$,\cite{hansen}
$
r^*(\omega)=r(-\omega)
$.
FDT holds for the bath:
$
r(\omega)-r(-\omega)= - i \omega \nu(\omega)
$.
In order to prove Eq.~(\ref{gg}),
we introduce a set of auxiliary variables $y_j$ 
satisfying the ordinary Langevin equation:
\begin{equation}
\left[ m_j \frac{d^2}{dt^2} + \Gamma_j \frac{d}{dt} + \Omega_j \right] y_j = \xi_j(t)- 
\frac{\partial F \left( \sum_j A_j y_j \right) }{\partial y_j}
\label{truc}
\end{equation}
with
$
\langle \xi_i(t) \xi_j(t_w) \rangle = 2 T^* \Gamma_j \delta_{ij} \delta(t-t_w) 
$.
The aim is to show that  $\sum_j A_j y_j$,
satisfies the same equations as $h$, Eqs.~(\ref{cincoff}) and (\ref{cincof}).\cite{footnote2}
The introduction of $y_i$ allows us to
convert the original non-Markovian problem 
into a Markovian one.
If $F$ is such that Eq.~(\ref{truc}) drives
the ensemble $y_i$ to equilibrium,  
\begin{equation}
P({\vec y}) \propto e^{
-\beta^*\left( \frac{\Omega_j y_j^2}{2 } + F\left(\sum_j A_j y_j \right) \right)
}
\; .
\label{dist-y}
\end{equation}

In Fourier space, Eq.~(\ref{truc}) reads
\begin{eqnarray*}
\hspace{-0.25cm}
[ - m_j (\omega-\omega_j^+)(\omega-\omega_j^-) 
+
\frac{\partial F(\sum_j A_j y_j) }{\partial y_j}
]
y_j(\omega)
&=&
\xi_j(\omega)
\end{eqnarray*}
where $\omega_j^{\pm}$ are the roots of 
$
-m_j \omega^{\pm \; 2} - i \omega^{\pm} \Gamma_j  + \Omega_j =0
$.
Defining 
$
h(\omega) = \sum_j A_j y_j(\omega)
$
one has
\begin{eqnarray}
h(\omega) &=&  
- \sum_j 
\frac{A_j \xi_j(\omega)+ A^2_j \partial_h F(h)}
{m_j (\omega-\omega_j^+)(\omega-\omega_j^-)} 
\label{langh1}
\; . 
\end{eqnarray}
Since $\mbox{\mbox Im} (\omega_j^{\pm}) \leq 0$,
we can choose the variables $y_j$ and the parameters
$m_j,\Gamma_j,\Omega_j$ in such a way to identify
\begin{equation}
r(\omega) = \sum_j \frac{ A^2_j}{m_j (\omega-\omega_j^+)(\omega-\omega_j^-)} 
\; .
\end{equation}
Furthermore, we associate the first term in the rhs  
of Eq.~(\ref{langh1}) to the coloured noise $\rho(\omega)$
and we use Eq.~(\ref{truc}) to obtain the noise-noise correlation:
\begin{eqnarray}
& & 
\langle \rho(\omega) \rho(\omega') \rangle =  2 T^* \delta(\omega+\omega') 
\nonumber\\
& & 
\times
\sum_j \frac{ \Gamma_j A^2_j}{m^2_j (\omega-\omega_j^+)(\omega-\omega_j^-)
(\omega+\omega_j^+)(\omega+\omega_j^-)}
\; .
\end{eqnarray}
The properties of the poles $(\omega_j^{\pm})^*=-\omega_j^{\mp}$ and
$m_j(\omega_j^++\omega_j^-)=-i\Gamma_j$ yield FDT
$
\nu(\omega) = -\frac{1}{i\omega} [ r(\omega)-r(-\omega)] 
$.

We need  an expression for $\Gamma_1\equiv\nu(t=0) =
\int_{-\infty}^\infty d\omega/(2\pi) \, \nu(\omega)$.
Completing the integration over the upper or the lower half plane, 
we have
\begin{eqnarray}
\Gamma_1&=&  
\sum_j \frac{A_j^2}{\Omega_j}
\label{nucero}
\end{eqnarray}

We have shown that Eq.~(\ref{cinco}) is equivalent to an ordinary
Markovian equation for ${\vec y}$, provided one identifies $h=\sum A_j y_j$.
The  stationary distribution for the $y_j$ is then given by Eq.~(\ref{dist-y}) 
${\hat P}(h)$ is calculated by introducing a delta-function:
\begin{eqnarray}
{\hat P}(h) &\propto& \int d\lambda  \int\Pi_j dy_j 
e^{-\beta^*\left(\frac{\Omega_j y_j^2}{2} + F(h) \right)-
    i\beta^* \lambda \left( \sum A_j y_j-h \right)} \nonumber \\
 &\propto& \int d\lambda 
e^{-\beta^*\left(\sum\frac{ A^2_j}{2\Omega_j} \lambda^2 - i \lambda h + F(h)
\right)} \nonumber \\
&\propto&  
e^{-\beta^*\left(\frac{h^2}{2 \Gamma_1} + F(h) \right)}
\; .
\label{nub}
\end{eqnarray}

The reciprocity property is immediate from the fact that, for a Markovian Langevin process
in equilibrium,
\begin{equation}
P({\vec y},t;{\vec y'},t_w)=P({\vec y'},t;{\vec y},t_w)
\; .
\end{equation}
 Similarly FDT follows from the fact that
 because  it holds
for any two functions of the $y_j$, it holds in particular for any two functions
of $h$.

\section{A Montecarlo algorithm with several baths}
\label{app.Monte}

We describe  the algorithm used to simulate the 
evolution of a system interacting with several thermal baths
of different type. 
Let us consider a dynamic spin $s=\pm 1$ with energy   
$E(s)$ that is in contact with 
a `fast' bath of inverse temperature 
$\beta$ and a `slow' 
bath of inverse temperature $\beta^*$ and time-scale $\tau^*$.
The generalization to several variables and many baths is straightforward.

The algorithm is as follows. 

{\it i}
With probability $1/\tau^*$ we generate a Gaussian random variable  $h$ with
$s$-dependent mean and variance:
\begin{equation}
\langle h \rangle = \beta^* \Gamma_1^2 s \; , \;\;\;\;\;\;\;\;\;\;\;
\langle h^2 \rangle - \langle h \rangle^2 = \Gamma_1^2
\; .
\end{equation}
$\Gamma_1$ is the strength  of the bath. 

{\it ii}
With probability $1-1/\tau^*$ we make ordinary Montecarlo updates of $s$ at temperature $1/\beta$, 
with an energy
$
E_{\sc tot}=E(s) - h  s
$.

We justify this algorithm as follows.
We model the slow bath by a system of ${\cal N}$ non-interacting
spins $\sigma_a=\pm 1$, $a=1,\dots,{\cal N}$, weakly 
coupled to $s$. The total energy of system and bath is given by
\begin{equation}
E_{\sc tot}= E(s) - \frac{\Gamma_1}{{\cal N}^{1/2}} s \sum_a \sigma_a = E(s) - h  s
\label{energytotal}
\; ,
\end{equation}
with $E(s)$ the energy of the free spin, $\Gamma_1$ the coupling constant and 
the `field' $h$ defined as
$
h \equiv   \Gamma_1/{\cal N}^{1/2}\, \sum_a \sigma_a
$.

We propose the following dynamics to the coupled system:

{\it i}
Frequent updates of $s$ keeping $\sigma_a$ fixed. The spin is then in presence of a
constant external field $h$.
These updates are done with any Montecarlo procedure, at  inverse temperature $\beta$.

{\it ii}
Unfrequent updates of all $\sigma_a$, i.e. of the field $h$,  keeping $s$ fixed. 
This evolution is controlled by an inverse  temperature $\beta^*$. 

If the sequence is long, one can assume that each $\sigma_a$ is in thermal equilibrium, 
at temperature $1/\beta^*$ with an energy 
$E(\sigma_a) \sim - \Gamma_1/{\cal N}^{1/2} s  \sigma_a$. Each configuration $\sigma_a=\pm 1$ 
has probabilities $p_+$ and $p_-$ given by:
\begin{equation}
p_{\pm}= 
\frac{ \exp(\pm \beta^* \frac{\Gamma_1}{{\cal N}^{1/2}} s ) } 
         {2 \cosh \left(\beta^* \frac{\Gamma_1}{{\cal N}^{1/2}} s \right) }
\; .
\end{equation}
The $\sigma_a$ are then identically distributed independent random variables.
It follows that, in the limit of large ${\cal N}$,  $h$ is a Gaussian random variable.
Its mean and variance are given by:
\begin{eqnarray*}
\langle h \rangle &=&  \frac{\Gamma_1}{{\cal N}^{1/2}}\sum_a \langle \sigma_a \rangle  
= 
\beta^*\Gamma_1^2 s+ O\left(\frac{1}{{\cal N}^{1/2}} \right) 
\label{meanh}
\\
\langle h^2 \rangle -\langle h \rangle ^2 &=&  \frac{\Gamma_1^2}{{\cal N}}  
\sum_a  \left( \langle \sigma_a^2 \rangle -\langle \sigma_a \rangle ^2 \right) 
= 
\Gamma_1^2 + O\left(\frac{1}{{\cal N}^{1/2}} \right)
\label{varianceh}
\end{eqnarray*}

As a check of the consistency of this proceedure we
examine if detailed balance holds, as expected, in
the particular case of a variable coupled to 
two baths with equal temperatures $\beta=\beta^*$. 
The transition probability per unit time to go from the 
state $s_A$ to the state $s_B$ is
\begin{equation}
P(A \to B) =
\frac{\Gamma_1}{\sqrt{2 \pi}} \int dh \; 
e^{-\frac{(h-\beta \Gamma_1^2 s_A)^2}{2 \Gamma_1^2} }
\; P_h(A \to B) 
\label{uno}
\end{equation}
where $P_h(A \to B)$ is the transition probability given $h$. We assume 
that $P_h(A \to B)$ satisfies detailed balance
\begin{equation}
\frac{P_h(A \to B)}{ P_h(B \to A)}= e^{-\beta (V(s_B)-V(s_A) - h(s_B-s_A))}
\label{detailedh}
\end{equation}
and we then prove that this implies that the transition probability over a 
time interval $\tau$, $P(A \to B, \tau)$, does too. 
Inserting  Eq.~(\ref{detailedh}) into Eq.~(\ref{uno}) we have:
\begin{eqnarray}
& & 
P(A \to B) 
= 
\frac{\Gamma_1}{\sqrt{2 \pi}} \int dh \; 
e^{ -\frac{(h-\beta \Gamma_1^2 s_A)^2}{2 \Gamma_1^2} }
\; P_h(B \to A) \; 
\nonumber\\
& & 
\;\;\;\;\;\;\;\;\;\;\;\;\;\;\;\;
\times 
e^{ -\beta \left( V(s_B)-V(s_A) - h(s_B-s_A) \right) }
\nonumber \\
& & = e^{ -\beta (V(s_B)-V(s_A)) }P(B \to A)
\end{eqnarray}
i.e. detailed balance.


\begin{thebibliography}{99}
\bibitem{footnote3}
A phenomenological definition of `fictive temperatures' 
has been intensively used to fit experimental data of 
glasses. We refer the reader to Ref.~\cite{Cukupe}
for a comparison of the fictive temperatures to $T^{\sc eff}$.
\bibitem{footnote5}
Let us stress that we call  `temperature' a quantity that is directly measurable and 
is related to the zero-th law of thermodynamics.
\bibitem{footnote1} 
The process being non-Markovian, one has to specify not only the position $x'$ at
a previous time, but also the  distribution of all previous positions.  Here we consider
the stationary measure: $P(x,t;x',t_w)$ stands for the joint probability of  $(x,t;x',t_w)$
 integrated over all histories from $-\infty$ to $t$.
\bibitem{footnote6} ERRATUM:  the boson-fermion transformation
in Ref.~\cite{Cuku-stat} should have read $K(t_a) \to K(t_s) + \epsilon \overline \theta_a$, 
$\overline \theta_a \to \overline \theta_a + \overline \epsilon$, $\dot K(t_a) \theta_a 
\to \dot K(t_a) \theta_a + \epsilon T$.
\bibitem{footnote2}
Note that we use the set of auxiliary variables 
$y_i$ to represent the original variable $h$,  
unlike the usual situation in which  auxiliary variables represent the bath.
\end{thebibliography}
\end{document}